\input psfig

\documentstyle{article}

\begin{document}

\vspace{1 cm}
\LARGE
\centerline
{\bf Statistical Mechanics of Charged Polymers in}
\centerline{\bf  Electrolyte
Solutions:
 A Lattice Field Theory Approach}

\vspace{1.8 cm}
\large
\centerline
{Stefan Tsonchev and Rob D. Coalson}
\centerline
{Dept. of Chemistry,
Univ. of Pittsburgh}
\centerline
{Pittsburgh, PA 15260}

\vspace{0.6 cm}
\centerline
{Anthony Duncan}
\centerline
{Dept. of Physics, Univ. of Pittsburgh}
\centerline
{Pittsburgh, PA 15260}

\newpage
\vspace{4 cm}
\centerline{\bf Abstract}

The Lattice Field
Theory approach to the statistical mechanics of a classical
Coulomb gas [R.D. Coalson and A. Duncan,
J. Chem. Phys. {\bf97}, 5653 (1992)] is generalized 
to include charged polymer chains. Saddle-point analysis is done on the functional integral representing the partition function of the full system.
Mean-field level analysis requires extremization of a real-valued functional
which possesses a single minimum, thus guaranteeing a unique solution.
The full mean-field equations for such a coupled system are derived, as well as the leading (one-loop) fluctuation corrections. Two different numerical real-space lattice procedures are developed to implement the generalized theory; these are applied to the problem of a charged  polymer confined to a spherical cavity in an electrolyte solution. The results provide new insight into the physics of confined polyelectrolytes.
\newpage
\section{Introduction}

   In a recent series of papers,
  the application of lattice field theory (LFT) techniques 
\cite{LFT1,LFT2,LFT3,LFT4,LFT5}
 to systems of mobile polar ions interacting with immobile macroions  has led
 to a  calculational framework for extracting free energies,
ionic concentrations, electrostatic potentials, and other physical properties.
 This formalism, which relies on a Hubbard-Stratonovich transformation
 of a functional integral representation of the grand canonical partition function, has a number of antecedents in the literature \cite{khol,DunMaw,pod}.  In the generalization introduced
 in \cite{LFT1}, we were able to obtain accurate results at the mean-field level,
 {\em with precisely computable corrections indicating the level of accuracy of
 the mean-field results}, for systems of highly variable ionic concentration and
 macromolecular charge.
 Paper \cite{LFT1} focussed on interactions between charged spherical colloid
 particles (polyballs) in electrolyte solutions.  In a subsequent paper, a 
 liquid phase assembly of many interacting polyballs was studied \cite{LFT2}.
  The formalism of
 \cite{LFT1} has been extended to include effects of a spatially variable 
dielectric
 profile \cite{LFT3}, the finite size of the simple 
 mobile ions forming the electrolyte \cite{LFT4}, and the presence of mobile 
dipoles
 in the gas (solution) \cite{LFT5}.  The objective of the present paper is
 to extend the LFT approach to compute the statistical mechanics of a gas of
 mobile charged ions to a system in which mobile charged polymer chains are
 also present and interacting electrostatically with the mobile ions.

 In Section 2
 we briefly  review  the  functional integral representation
 (and lattice field theory discretization thereof) for the grand
 canonical partition function introduced in \cite{LFT1}.
 In Section 3 the formalism is extended to include charged polymer chains. The
 mean-field level solution of the full system (ions plus polymers) is then
 presented in Section 4. Our results are similar to, but do not fully agree with 
some recent
 work of Orland et al \cite{Orland}. In Section 5 we justify the contour 
deformation
 introduced in Section 4 to extract the leading saddle-point (or mean-field 
theory)
 equations of the full ion-polymer system by deriving a general convexity
 theorem for the mean-field free energy. An explicit expression is also derived 
for
 the leading post-mean-field corrections due to the fluctuations around the 
saddle-point
 fields. This means that a precise estimate of the error in the mean-field theory 
is
 in principle possible even in systems of coupled ions and charged polymer chains.
 Some simple illustrative  applications of the mean-field
 results of Section 4 to systems possessing spherical symmetry,
 for which the problem can be reduced to the minimization of a one-dimensional functional, are described in Section 6. The same systems are treated in Section 7 by solving the 
mean-field equations on a three-dimensional lattice, an approach which can be used to study systems of arbitrary shape and symmetry.
 Section 8 summarizes our results and indicates areas for further research. 

\newpage

\section{Functional formalism and lattice field theory for ionic systems}

  A simple functional integral representation for the grand canonical partition function
 of a system of mobile ions with electric charge density $\rho(\vec{r})$ is
 obtained by rewriting the electrostatic potential energy of the system via
 a standard functional identity:

\begin{equation}
e^{-\frac{\beta}{2\epsilon}\int 
d\vec{r}d\vec{r}\,^{\prime}\frac{\rho(\vec{r})\rho(\vec{r}\,^{\prime})}
{|\vec{r}-\vec{r}\,^{\prime}|}}=C\int D\chi 
e^{\frac{\epsilon}{8\pi\beta}\int\chi\Delta\chi
d\vec{r}+i\int\chi(\vec{r})\rho(\vec{r})d\vec{r}}
\end{equation}
Here $C$ is an irrelevant constant (independent of $\rho$) which will be neglected 
from now on, and the charges are immersed in
 a medium of dielectric constant $\epsilon$ \cite{footeps}
at inverse temperature $\beta = 1/kT$. For simplicity,
 assume the ions to be of two varieties only: $n_{+}$ ions of charge $+e$ at 
locations
 $\vec{r}\,^{+}_{k}$ and $n_{-}$ ions of charge $-e$ at locations 
$\vec{r}\,^{-}_{l}$
 (with $1\leq k\leq n_{+}, 1\leq l\leq n_{-}$). Of course, if there are no other charged entities present, $n_{+}=n_{-}$, but for the time being (anticipating the inclusion of charged polymer chains below) we leave $n_{\pm}$ free. The charge density is then
\begin{equation}
\rho(\vec{r})=e\sum_{k}\delta(\vec{r}-\vec{r}\,^{+}_{k})-e\sum_{l}\delta(\vec{r}-\vec{r}\,^{-}_{l})
\end{equation}

The effect of the transformation induced by (1) is twofold. First,
the nonlocal Coulomb potential (which makes the direct simulation of such
 systems by molecular dynamics extremely difficult) has been replaced by a local Laplacian operator. Secondly, the nonlinear pairwise coupling of the charged ions has been replaced by a linear coupling term where the ions       
interact only with the auxiliary field. Consequently, for a fixed 
$\chi$ field, the statistical mechanics of the ionic gas is 
elementary. Introducing chemical potentials $\mu_{\pm}$ for the positive and 
negative
 ions, the sum of the factor $\exp{(i\int\chi(\vec{r})\rho(\vec{r})d\vec{r})}$
 over ion numbers and positions can be explicitly performed:

\begin{eqnarray}
&&\sum_{n_{k},n_{l}}\frac{1}{n_{k}!}\frac{1}{n_{l}!}\int\prod\frac{d\vec{r}_{k}}{\lambda_{+}^3}
\frac{d\vec{r}_{l}}{\lambda_{-}^3} 
  e^{ie\sum_{k}\chi(\vec{r}_{k})-ie\sum_{l}\chi(\vec{r}_{l})+\beta(\mu_{+}n_{k}
 +\mu_{-}n_{l})} \nonumber \\
 &=& \exp{(e^{\beta\mu_{+}}\int\frac{d\vec{r}}{\lambda_{+}^3}e^{ie\chi}
 +e^{\beta\mu_{-}}\int\frac{d\vec{r}}{\lambda_{-}^3}e^{-ie\chi})}
\end{eqnarray}
where $\lambda_{\pm}$ are thermal deBroglie wavelengths for the ions. Inserting
 this result in the complete expression for the Boltzmann weight (1), the full
 grand canonical partition function becomes
\begin{equation}
\label{eq:contpart}
 Z_{\rm{gc}}=\int D\chi e^{\frac{\beta\epsilon}{8\pi}\int\chi\Delta\chi d\vec{r}
   +e^{\beta\mu_{+}}\int e^{ie\beta\chi}d\vec{r}/\lambda_{+}^{3}
   +e^{\beta\mu_{-}}\int e^{-ie\beta\chi}d\vec{r}/\lambda_{-}^{3}}
\end{equation}
 where we have rescaled the auxiliary field $\chi\rightarrow \beta\chi$.
 The generalization of this expression to include stationary charged macroions 
\cite{LFT1},
 external single particle potentials acting on the mobile ions, spatially varying 
dielectric
 constant \cite{LFT3}, finite size
 effects \cite{LFT4}, and effects due to mobile multipolar entities \cite{LFT5}
 have been discussed elsewhere and will not
 be treated here. For example, the inclusion of a single particle potential 
$V(\vec{r})$
 acting on the ions can be effected simply by the replacement $e^{\pm ie\beta\chi}
\rightarrow e^{\pm ie\beta\chi -V}$ in (\ref{eq:contpart}). Such a potential can 
be used 
 (for example) to exclude the ions from certain regions of space.

  The functional integration indicated in (\ref{eq:contpart})  may be made 
well-defined by 
discretizing the system on a cubical lattice defined by lattice points
$\vec{r}=a_{l}(n_x,n_y,n_z)$, where $a_{l}$ is a lattice spacing taken small in
comparison to the length scales over which the field $\chi$ varies substantially,
and $n_x,n_y,n_z$ are integers, $0\leq n_x,n_y,n_z \leq L-1$, for a $L$x$L$x$L$
lattice. We shall restrict our calculations to
 systems with zero total electric charge, where the fixed charges
are screened by the mobile ones and the fields fall exponentially far from the 
sources,
so finite volume effects go to zero rapidly as the lattice size $La_{l}$ is 
increased. 
 With these notations, the lattice
version of the transformed partition sum (\ref{eq:contpart}) becomes
\begin{eqnarray}
\label{eq:lattpart}
Z_{\rm gc} &=&  \int \prod_{\vec{n}} d\chi_{\vec{n}}
\exp \left\lbrace \frac {\alpha}{2}
 \sum_{\vec{n}\vec{m}} \chi_{\vec{n}} \Delta_{\vec{n}\vec{m}} 
\chi_{\vec{m}}  \right. \;  \nonumber \\ 
&+& \left. \gamma_+ \sum_{\vec{n}} e^{ie\beta\chi_{\vec{n}}} 
 \; +
\gamma_- \sum_{\vec{n}} e^{-ie\beta\chi_{\vec{n}}} 
  \right\rbrace    \;
\end{eqnarray}
Here $\Delta_{\vec{n}\vec{m}}$ is the discrete lattice Laplacian operator, which
may be regarded as a $L^3$x$L^3$ matrix  (see \cite{LFT1}). The dimensionless 
constants
$\alpha\equiv a_{l}\beta\epsilon/4\pi$, $\gamma_{\pm}\equiv 
e^{\beta\mu_{\pm}}(a_{l}/\lambda_{\pm})^3$
 have been introduced to simplify the notation.  In (5), $\Delta$ now
represents the discrete lattice gradient (nearest-neighbor difference operator),
with $e$ the electronic charge and  $a_{l}$ the lattice spacing. In Section 4 we 
shall return
 to the evaluation of the integral (5) by saddle-point techniques, after 
 the additional terms implementing charged polymer chains are included.

\newpage

\section{Functional formalism for charged polymer chains}

  The path integral representation of a polymer chain \cite{Edwards,wieg}
 is well-known.
 We shall show below that for a system consisting of many monomers the leading
 behavior is determined simply by the total length $M$ (= total number of 
monomers) of all chains 
 and not by the connectivity or number of the separate chains. Thus the 
configuration
 of the polymer chain(s) can be represented by a single function $\vec{x}(s)$, 
where
 the dimensionless path length variable $0\leq s \leq M$. If the typical monomer
 separation is $a_p$ (the Kuhn length \cite{Edwards}), the partition sum for
 such a system may be written
\begin{eqnarray}
    Z_{{\rm pol}}&=&\int D\vec{x}(s)e^{-\frac{3}{2a_{p}^{2}}\int_{0}^{M}ds 
\dot{\vec{x}}^{2}(s)
 -\frac{\lambda}{2}\int \phi_{p}(\vec{r})^{2}d\vec{r}}  \\
    \phi_{p}(\vec{r}) &=& \int ds \delta(\vec{r}-\vec{x}(s))
\end{eqnarray}
 The field $\phi_p$ is the spatial monomer density and the term involving 
$\lambda\int
\phi_{p}^{2}d\vec{r}$ prevents the monomers from overlapping with each other,
$\lambda \ge 0$ being a measure of monomer excluded volume.
 As discussed at the end
of Section 4, $\lambda \simeq B_2 \simeq \sigma^3$, where $B_2$ is the second 
virial
coefficient, which goes approximately like the monomer volume (i.e. $\sigma$ is
the effective hard-sphere radius of a monomer) \cite{McQ}.
 For simplicity we shall assume
 initially that the polymer charge density is uniform along the polymer chains, 
with each
 monomer carrying a net charge of  $-pe$ \cite{Orland} (the generalization to 
varying
 monomer charge would simply result in an effective Schr\"odinger Hamiltonian with
 a time-dependent potential (cf. discussion below)).  The negative sign indicates 
that we have taken the
 polymer chain to be negatively charged, as a consequence of dissociation of
 positively charged counterions. This means that (2) for the total charge density 
now
 becomes \cite{footmac}
\begin{equation}
\rho(\vec{r})=e\sum_{k}\delta(\vec{r}-\vec{r}\,^{+}_{k})-e\sum_{l}\delta(\vec{r}-\vec{r}\,^{-}_{l})
   -pe\phi_{p}(\vec{r})
\end{equation}

 Repeating the argument leading from (2) to (4) we now find the following 
expression for
 the full grand canonical partition function of the system
\begin{eqnarray}
 Z_{\rm{gc}}&=&\int D\chi 
D\phi_{p}D\vec{x}(s)e^{\frac{\beta\epsilon}{8\pi}\int\chi\Delta\chi d\vec{r}
     -\frac{3}{2a_{p}^{2}}\int_{0}^{M}ds 
\dot{\vec{x}}^{2}(s)-\frac{\lambda}{2}\int\phi_{p}(\vec{r})^{2}d\vec{r}} \nonumber \\
  &\times& e^{e^{\beta\mu_{+}}\int e^{ie\beta\chi}d\vec{r}/\lambda_{+}^{3}
         +e^{\beta\mu_{-}}\int e^{-ie\beta\chi}d\vec{r}/\lambda_{-}^{3} 
-ipe\beta\int\chi(\vec{r})\phi_{p}(\vec{r})d\vec{r}}
\end{eqnarray}
 The integration over polymer configurations can now be replaced by an equivalent
 Schr\"odinger Hamiltonian problem. To facilitate this, linearize the dependence
 on the monomer density $\phi_{p}$ by introducing a second auxiliary field
\begin{equation}
  e^{-\frac{\lambda}{2}\int\phi_{p}^{2}d\vec{r}}=\int D\omega 
e^{-\frac{\lambda}{2}\int \omega^{2}
d\vec{r}-i\lambda\int\omega(\vec{r})\phi_{p}(\vec{r})d\vec{r}}
\end{equation}
Introducing this representation into (9) and using (7), one finds
\begin{equation}
 Z_{\rm{gc}}=\int 
D\chi(\vec{r})D\omega(\vec{r})e^{\frac{\beta\epsilon}{8\pi}\int\chi\Delta\chi 
d\vec{r}-\frac{\lambda}{2}\int\omega(\vec{r})^{2}d\vec{r}+c_{+}\int 
e^{ie\beta\chi}d\vec{r}+c_{-}\int e^{-ie\beta\chi}d\vec{r}}Z_{{\rm 
Schr}}(\chi,\omega)
\end{equation}
with $c_{\pm}=e^{\beta\mu_{\pm}}/\lambda_{\pm}^{3}$, and 
\begin{equation}
Z_{{\rm Schr}}(\chi,\omega)\equiv \int 
D\vec{x}(s)e^{-\frac{3}{2a_{p}^{2}}\int_{0}^{M}ds 
\dot{\vec{x}}^{2}(s)-ipe\beta\int  ds\chi(\vec{x}(s))-i\lambda\int 
ds\omega(\vec{x}(s))}
\end{equation}
 defines a functional which is just the Feynman path integral for the imaginary 
time 
 evolution of a particle of mass $3/a_{p}^{2}$ in an imaginary potential 
$i(pe\beta\chi+\lambda\omega)$. In fact, we shall  show  below that the functional 
integral (11)
 over $\chi$ and $\omega$ can be rerouted through a complex saddle-point at
 $\chi = -i\chi_{c}$ and $\omega= -i\omega_{c}$ where $\chi_{c}$ and $\omega_{c}$
 are purely {\em real}, so that the evaluation of $Z_{{\rm Schr}}(\chi,\omega) $ 
at the
 saddle reduces to a completely conventional three-dimensional
 Schr\"odinger Hamiltonian problem---namely,
 the computation of matrix elements of $e^{-HT}$ where the Euclidean time extent of the
 evolution is just $T=M$ and
\begin{equation}
  H\equiv -\frac{a_{p}^{2}}{6}\vec{\nabla}^{2}+\lambda\omega_{c}(\vec{r})
+\beta pe\chi_{c}(\vec{r})
\label{eq:H}
\end{equation}
For large $M$, any matrix element of  $e^{-HT}$ has the asymptotic form
\begin{eqnarray*}
\log{\left(\left<...\left|e^{-HM}\right|...\right>\right)}\simeq -ME_{0}+{\rm O(1)}
\end{eqnarray*}
 where $E_{0}$ is the ground state eigenvalue
 of the Hamiltonian $H$ \cite{footground}.
 Different boundary  conditions (periodic, open, etc) imposed at
 the ends of the polymer chain (or even a finite but fixed number of splits in the 
chain)
 correspond to the choice of different initial and final states in the above 
formula,
 contribute to the O(1) edge correction, and are subdominant for large $M$.

\newpage
\section{Mean-field (saddle-point approximation) theory of  coupled ionic-polymer 
system}
  The equations determining the saddle-point configuration fields 
$\chi_{c},\omega_{c}$
 are obtained by setting the variational derivative of the exponent in the full functional
 integral (11) to zero. After rotating the fields to the negative imaginary axis 
(the need
 for this rotation will be justified below when we discuss the fluctuation 
corrections), this exponent
 becomes
\begin{equation}
  F=\int d\vec{r}\left\lbrace 
\frac{\beta\epsilon}{8\pi}\left|\vec{\nabla}\chi_{c}\right|^{2}+\frac{\lambda}{2}
\omega_{c}^{2}+c_{+} e^{\beta e\chi_{c}}+c_{-} e^{-\beta e\chi_{c}}\right\rbrace
-ME_{0}(\chi_{c},\omega_{c})
\end{equation}
The functional derivatives determining the saddle-point solution are then
\begin{eqnarray}
    \frac{\delta E_{0}}{\delta\chi_{c}(\vec{r})} &=& \beta 
pe|\Psi_{0}(\vec{r})|^{2} \\
    \frac{\delta E_{0}}{\delta\omega_{c}(\vec{r})} &=& 
\lambda|\Psi_{0}(\vec{r})|^{2}\\
    \frac{1}{\beta e}\frac{\delta F}{\delta\chi_{c}(\vec{r})} &=& 
-\frac{\epsilon}{4\pi e}\vec{\nabla}^{2}\chi_{c}(\vec{r})+ c_{+}e^{\beta 
e\chi_{c}(\vec{r})}
 - c_{-}e^{-\beta e\chi_{c}(\vec{r})}-M p|\Psi_{0}(\vec{r})|^{2}=0 \\
   \frac{1}{\lambda} \frac{\delta F}{\delta\omega_{c}(\vec{r})} 
&=&\omega_{c}(\vec{r})- M
|\Psi_{0}(\vec{r})|^{2}=0
\end{eqnarray}
Here the wavefunction $\Psi_{0}(\vec{r})$ is assumed unit-normalized:
\begin{equation}
  \int |\Psi_{0}(\vec{r})|^{2}d\vec{r}=1
\end{equation}
Using (18), the auxiliary field $\omega_{c}$ may be eliminated completely, leaving the
 pair of coupled nonlinear equations as the complete mean-field solution to the full interacting ion-charged polymer problem:
\begin{eqnarray}
  \frac{\epsilon}{4\pi e}\vec{\nabla}^{2}\chi_{c}(\vec{r})&=&c_{+}e^{\beta 
e\chi_{c}(\vec{r})}
 - c_{-}e^{-\beta e\chi_{c}(\vec{r})}-Mp|\Psi_{0}(\vec{r})|^{2}
\label{eq:pb}   \\
  \frac{a_{p}^{2}}{6}\vec{\nabla}^{2}\Psi_{0}(\vec{r})&=&\lambda 
M|\Psi_{0}|^{2}\Psi_{0}(\vec{r})
+\beta pe\chi_{c}(\vec{r})\Psi_{0}(\vec{r})-E_{0}\Psi_{0}(\vec{r})
\label{eq:nlse}
\end{eqnarray}
 As mentioned previously, the inclusion of single particle potentials, which can be used to  enforce exclusion regions for either the ions or the monomers, is straightforward.
 Using the notation $V_{c}(\vec{r})$ (resp. $V_{m}(\vec{r})$) for potential fields acting on the ions (resp. monomers) (20--21) become
\begin{eqnarray}
  \frac{\epsilon}{4\pi e}\vec{\nabla}^{2}\chi_{c}(\vec{r})&=&c_{+}e^{\beta 
e\chi_{c}(\vec{r})-V_{c}(\vec{r})}
 - c_{-}e^{-\beta e\chi_{c}(\vec{r})-V_{c}(\vec{r})}-Mp|\Psi_{0}(\vec{r})|^{2}\\
  \frac{a_{p}^{2}}{6}\vec{\nabla}^{2}\Psi_{0}(\vec{r})&=&\lambda 
M|\Psi_{0}|^{2}\Psi_{0}(\vec{r})
+\beta 
pe\chi_{c}(\vec{r})\Psi_{0}(\vec{r})-(E_{0}-V_{m}(\vec{r}))\Psi_{0}(\vec{r})
\end{eqnarray}

  Recalling that the parameters $c_{\pm}$ are exponentials of the chemical 
potentials 
 $\mu_{\pm}$ for positively and negatively charged ions, the numbers of
 these ions must be fixed by suitably adjusting $c_{\pm}$, using the easily 
derived
 relations:
\begin{equation}
  n_{\pm}=c_{\pm}\frac{\partial \log{(Z_{\rm gc})}}{\partial c_{\pm}}=c_{\pm}
\int e^{\pm\beta e\chi_{c}-V_{c}}d\vec{r}
\end{equation}
 while the number density of monomers is given by $M|\Psi_{0}(\vec{r})|^{2}$ which clearly  integrates to the total number of monomers $M$.  Charge neutrality follows
 from integrating (22) over all space (correct either for periodic boundary 
conditions or for an electric field falling to zero at the system boundaries):
\begin{eqnarray}
  0&=&\int \vec{\nabla}^{2}\chi_{c}(\vec{r})d\vec{r}\Rightarrow 0=n_{+}-n_{-}-Mp 
\\
  &\Rightarrow& n_{+}e = n_{-}e + Mpe
\end{eqnarray}

  For systems of spherical symmetry, the equations (22--23) reduce to a pair of coupled nonlinear ODEs:
\begin{eqnarray}
    \eta\frac{d^{2}F}{dr^{2}}&=& 
r\left(\xi_{+}e^{F(r)/r-V_{c}(r)}-\xi_{-}e^{-F(r)/r-V_{c}(r)}\right)-\frac{1}{r}G(r)^{2} \\
\frac{1}{6}\frac{d^{2}G}{dr^{2}}&=&\frac{p}{r}F(r)G(r)+\frac{1}{r^{2}}G(r)^{3}-(E_
{0}-V_{m}(r))G(r)
\end{eqnarray}
 where the rescaled radial functions $F$, $G$ are defined as
\begin{eqnarray}
  F(r) &=& \beta e r\chi_{c}(r) \\
  G(r) &=& \sqrt{\lambda M}r\Psi_{0}(r)
\end{eqnarray}
 The radial variable in (27--28) is measured in units of the monomer size $a_{p}$, 
and we have defined dimensionless variables
\begin{eqnarray}
  \eta &=& \frac{\lambda\epsilon}{4\pi \beta pe^{2}a_{p}^{2}} \\
  \xi_{\pm} &=& \frac{\lambda c_{\pm}}{p}
\end{eqnarray}
In terms of these new variables the charge and number density constraints (24--26)
 become
\begin{eqnarray}
  \xi_{\pm}\int r^{2}e^{\pm \frac{F(r)}{r}-V_{c}(r)}dr &=& \frac{\lambda}{4\pi pa_{p}^{3}} n_{\pm}\\
  \int G(r)^{2}dr &=& \frac{\lambda}{4\pi a_{p}^{3}}M  \nonumber  \\
                           &=&\frac{\lambda}{4\pi a_{p}^{3}p}(n_{+}-n_{-})
\end{eqnarray}

  In fact, we shall show below that the solution of the coupled nonlinear 
equations (27--28) is
 most readily accomplished by returning to the free energy expression (14), which 
represents a
 functional of two scalar fields $\chi_{c}$, $\omega_{c}$. This functional will be 
shown to be
 convex and bounded below, with a unique minimum at the field values satisfying 
(27--28).

  For non-spherically-symmetric systems, a practical approach  is again
provided by the lattice field
 theory discretization of \cite{LFT1}. The inclusion of the polymer terms in (5) 
is quite straightforward---discretizing (11), we obtain (setting the single particle potentials
$V_{c}$, $V_{m}$ to zero for simplicity)
\begin{eqnarray}
\label{eq:lattpartfull}
Z_{\rm gc} &=&  \int \prod_{\vec{n}} d\chi_{\vec{n}}d\omega_{\vec{n}}
\exp \left\lbrace \frac {\alpha}{2}
 \sum_{\vec{n}\vec{m}} \chi_{\vec{n}} \Delta_{\vec{n}\vec{m}} 
\chi_{\vec{m}}  -\frac{\lambda}{2}a_{l}^{3}\sum_{\vec{n}}\omega_{\vec{n}}^{2}
\right. \;  
\nonumber \\ 
&+& \left. \gamma_+ \sum_{\vec{n}} e^{ie\beta\chi_{\vec{n}}} 
 \; +
\gamma_- \sum_{\vec{n}} e^{-ie\beta\chi_{\vec{n}}} 
 -ME_{0}(\chi_{\vec{n}},\omega_{\vec{n}}) \right\rbrace    \;
\end{eqnarray}
  where $E_{0}$ is the ground state energy of the associated discrete Hamiltonian matrix
\begin{equation}  
H_{\vec{m}\vec{n}}=-\frac{a_{p}^{2}}{6a_{l}^{2}}\Delta_{\vec{m}\vec{n}}
+i\lambda\omega_{\vec{n}}
 +i\beta pe\chi_{\vec{n}}
\end{equation}
 Of course, at the complex saddle-point of the integral (35), $\chi = -i\chi_{c}$ 
and
 $\omega = -i\omega_{c}$ with $\chi_{c}$, $\omega_{c}$ real fields, so the 
Hamiltonian 
 matrix (36) is real symmetric, with well-defined real ground state energy. After 
the rotation
 to the imaginary axis, we find the following expression for the discretized free energy
 functional (corresponding to the continuum expression given in (14)):
\begin{equation}
  F= 
-\frac{\alpha}{2}\sum_{\vec{m}\vec{n}}\chi_{\vec{m}}\Delta_{\vec{m}\vec{n}}\chi_{\vec{n}}
 +\frac{\lambda}{2}a_{l}^{3}\sum_{\vec{n}}\omega_{\vec{n}}^{2} 
+\gamma_{+}\sum_{\vec{n}}
e^{\beta e\chi_{\vec{n}}}+\gamma_{-}\sum_{\vec{n}}e^{-\beta e 
\chi_{\vec{n}}}-ME_{0}(\chi_{\vec{n}},\omega_{\vec{n}})
\end{equation}
where we have dropped the subscript $c$ indicating the saddle-point (mean-field) value for the fields to avoid notational overload. Now the ground state energy, as promised, is
 a perfectly real number, namely the lowest eigenvalue of the matrix
\begin{equation}  
H_{\vec{m}\vec{n}}=-\frac{a_{p}^{2}}{6a_{l}^{2}}\Delta_{\vec{m}\vec{n}}+\lambda
\omega_{\vec{n}}
 +\beta pe\chi_{\vec{n}}
\end{equation}

For a long Gaussian polymer chain in a confined region and subjected to
an {\it external} potential $W (\vec r)$, it is well-known \cite{Edwards}
that the ground state of the three dimensional
Schr\"odinger operator
\begin{equation}
\hat H  
  = -\frac{a_{p}^{2}}{6}\vec{\nabla}^{2}+ \beta W (\vec r)
\label{eq:W}
\end{equation}
determines the equilibrium properties of the polymer chain.  In particular,
if $\Psi_0 (\vec r)$ is the unit-normed ground state eigenfunction
associated with Hamiltonian (\ref{eq:W}) (with $\Psi_0 =0$ on
the confining boundary surface) then $|\Psi_0 (\vec r)|^2$
is the normalized monomer density.  Equivalently, if there are $M$ monomers in the 
polymer system then $M|\Psi_0 (\vec r)|^2$ is the number density of monomers in 
the system.  Comparing (\ref{eq:W}) to (38), (18) above,
we see that in our problem there is an {\it effective} potential
$$W(\vec r) = kT \lambda M |\Psi_0 (\vec{r})|^2 + pe\chi_c (\vec r)$$
Since both terms in this potential are functions (explicitly or
implicitly) of $\Psi_0$, our effective potential $W$ is evidently a 
{\it mean-field} potential.  Specifically, the electric potential
$\chi_c$ depends on the monomer charge density (hence $\Psi_0$) according to (20). 
  The other contribution to the effective potential, i.e. $kT \lambda M |\Psi_0 (\vec{r})|^2$, is due to short-range excluded volume interactions between 
monomers. Its origin and form can be understood as follows. Let $U(\vec r)$
be a short range repulsive potential via which all monomer pairs interact.
Then, if there is a distribution $n_m (\vec r)$ of monomers in the system,
the repulsive potential (call this $W$, suppressing the electrostatic
contribution) experienced by a test monomer inserted
at point $\vec r$ is:
\begin{eqnarray}
W(\vec r) &=& \int d \vec r \, ' U(\vec r- \vec r \, ') n_m (\vec r \, ')\\
     &\cong& n_m (\vec r) \int d\vec r \,U(\vec r)  
\end{eqnarray}
The second line follows from the first under the assumption
that the pair potential $U$ is short range compared to the
length scale on which the monomer density $n_m$ varies. Using
the connection (noted above) that $M |\Psi_0 (\vec r)|^2 = n_m (\vec r)$,
we immediately identify the parameter $\lambda$ as
\begin{equation}
\lambda  = \beta  \int U(\vec r) d\vec r 
\end{equation}
This can in turn be connected to the second virial coefficient $B_2$,
which can be used to ascribe an effective ``hard sphere" radius $\sigma$
to the monomer \cite{McQ}:
\begin{equation}
 \beta  \int U(\vec r) d\vec r  \cong
- \int d\vec r \left[ e^{-\beta U(\vec r)} - 1 \right]  = 
2 B_2 \cong 4 \pi \sigma^3 /3
\end{equation}
In this way we can connect $\lambda$ with the effective size
of a monomer.

\newpage
\section{Corrections to mean-field theory---convexity of the free energy}

  In the preceding section we claimed that the functional integral (11) is 
dominated by a 
 complex saddle-point where the fields $\chi,\omega$ take pure imaginary values  $-i\chi_{c},-i\omega_{c}$. To justify this assertion, we need to ensure that the deformation
 of  the contour of the field integrations is such that the integral passes 
through a 
 proper saddle-point, with a maximum of the real part of the exponent, and that 
the saddle
 chosen is the dominant one globally.  A rigorous discussion presupposes that the 
functional
 integral (11) has been made well-defined, say by discretization as in (35).
 As a consequence we
 may be sure that the contour deformation is possible in the first
 place as a result of the analyticity of  $Z_{\rm Schr}={\rm 
Tr}e^{-MH(\chi_{\vec{n}},\omega_{\vec{n}})}$
 as a function of the variables $\chi_{\vec{n}}$, $\omega_{\vec{n}}$ for an 
arbitrary 
 finite-dimensional matrix $H$. We remind the reader that the result $Z_{\rm Schr}
 ={\rm Tr}e^{-MH}$ holds for closed polymer chains but that different boundary conditions
 induce subdominant edge effects of order O(1/M) for long chains with M monomers 
(cf. discussion following (13)). The
 analyticity of the remaining part of the integrand in (35) is obvious.  
   
   For simplicity we shall temporarily return to continuum notation, although the
 reader is warned that ultimately the functional integral being discussed must be
 explicitly defined by a cutoff procedure. Writing
\begin{eqnarray}
   \chi(\vec{r}) = -i\chi_{c}(\vec{r})+\hat{\chi}(\vec{r})  \nonumber \\
   \omega(\vec{r}) = -i\omega_{c}(\vec{r})+\hat{\omega}(\vec{r})
\end{eqnarray}
 we may expand the integrand of the functional integral (11) keeping terms up to 
second
order in the fluctuation fields $\hat{\chi}$, $\hat{\omega}$:
\begin{eqnarray}
 Z_{\rm gc}&= &e^{F}\int D\hat{\chi}(\vec{r})D\hat{\omega}(\vec{r})
e^{-\int 
d\vec{r}\left(\frac{\beta\epsilon}{8\pi}\left|\vec{\nabla}\hat{\chi}(\vec{r})\right|^{2}+\frac{\lambda}{2}\hat{\omega}(\vec{r})^{2}
+\frac{\beta^{2}e^{2}}{2}\left(c_{+}e^{\beta e\chi_{c}}
 +c_{-}e^{-\beta e\chi_{c}}\right)\hat{\chi}(\vec{r})^{2}\right)} \nonumber \\
 &\times&e^{-M\int d\vec{r}d\vec{r}\,^{\prime}(\lambda\hat{\omega}(\vec{r})+
\beta pe\hat{\chi}(\vec{r}))G_{c}(\vec{r},\vec{r}\,^{\prime})(\lambda\hat{\omega}(\vec{r}\,^{\prime})+\beta pe\hat{\chi}(\vec{r}\,^{\prime}))}
\end{eqnarray}
where the prefactor $e^{F}$ contains the entire mean-field result (14) and the ``one-loop"
fluctuation corrections to mean-field are contained in the remaining Gaussian 
functional
 integral over the fluctuation fields $\hat{\chi}$, $\hat{\omega}$. The crucial point is that this
 integral is convergent as a consequence of the positivity of the Green's function
 $G_{c}(\vec{r},\vec{r}\,^{\prime})$ giving the second order variation with 
$\hat{\chi}$, $\hat{\omega}$ of the ground state 
 eigenvalue $E_{0}$ of the hermitian Hamiltonian (13). Let $\Psi_{n}(\vec{r})$ be 
a complete
 orthonormal set of eigenfunctions of $H$, with the corresponding 
 ordered eigenvalues $E_{n}$, $E_{n}\leq E_{m}$ for $n<m$. Then standard second order
 perturbation theory  gives, as a consequence of (44),
\begin{eqnarray}
 E_{0}(\chi,\omega)&\simeq&E_{0}(\chi_{c},\omega_{c})+\int 
d\vec{r}d\vec{r}\,^{\prime}(\lambda\hat{\omega}(\vec{r})+\beta 
pe\hat{\chi}(\vec{r}))G_{c}(\vec{r},\vec{r}\,^{\prime})(\lambda\hat{\omega}(\vec{r}\,^{\prime})+\beta pe\hat{\chi}(\vec{r}\,^{\prime})) \nonumber \\
G_{c}(\vec{r},\vec{r}\,^{\prime})&=&\Psi_{0}(\vec{r})\Psi_{0}(\vec{r}\,^{\prime})\sum_{n\neq 0}
\frac{\Psi_{n}(\vec{r})\Psi_{n}(\vec{r}\,^{\prime})}{E_{n}-E_{0}}
\label{eq:G}
\end{eqnarray}
through terms of second order in the fluctuation fields.
Note that the first order perturbation shift in $E_{0}$ is canceled at the 
saddle-point by the
first order variation in the exponent in (11).
Furthermore,  the positivity (strictly speaking, positive semidefiniteness) of $G_{c}$ is manifest.
To see this, note that a necessary and sufficient condition for $G_{c}$ to be 
positive semidefinite is
\begin{equation}
\int d \vec r \int d \vec r \, ' f(\vec r) G_c (\vec r, \vec r \, ' ) f(\vec r \, ') \ge 0
\label{eq:pd}
\end{equation}
for any function $f( \vec r)$.  Using the definition given in (\ref{eq:G}),
\begin{equation}
\int d \vec r \int d \vec r \, ' f(\vec r) G_c (\vec r, \vec r \, ' ) f(\vec r \, ') =
\sum_{n \ne 0} \frac{ \left[ \int d\vec r \,\Psi_0 (\vec r) f(\vec r) \Psi_n (\vec r) \right]^2}
{E_{n}-E_{0}}
\end{equation}
Since the denominator of each term on the r.h.s. is positive and the numerator
is non-negative the condition for positive semidefiniteness (\ref{eq:pd}) is 
satisfied.

  It should be emphasized that the positivity of the full fluctuation kernel in (45) holds for arbitrary
 real fields $\chi_{c},\omega_{c}$---it is not essential that they satisfy the saddle-point equations
 (20--21) (although we shall of course eventually demand that they do!). The value of this
 observation is that it is equivalent to a statement of  convexity of the free energy functional
 $F(\chi_{c},\omega_{c})$ in (14) for arbitrary values of its field arguments, as the exponent
 in the fluctuation integral (45) is essentially the second functional derivative of $F$. The
 functional $F(\chi_{c},\omega_{c})$ is clearly bounded below (as the lowest 
eigenvalue
 of $H$ in (13) can grow at most linearly with $\chi_{c}$ or $\omega_{c}$). If it is everywhere
 convex, it must have a {\em unique} minimum at exactly the field values 
$\chi_{c},\omega_{c}$
 satisfying (20--21). Thus the stable minimization procedures employed in 
\cite{LFT1} for
 solving the lattice field theory of systems of polyballs and mobile ions are 
guaranteed to work
 here also, in the presence of long charged polymer chains, provided an efficient numerical technique is employed!

 We now return to the task of evaluating the Gaussian fluctuation integral (45). It is 
 convenient to change the field integration variables by replacing the excluded volume field
 $\hat{\omega}(\vec{r})$ with the linear combination
\begin{equation}
   \sigma(\vec{r}) \equiv \lambda\hat{\omega}(\vec{r})+\beta p 
e\hat{\chi}(\vec{r})
\end{equation}
so that the fluctuation integral in (45) becomes
\begin{eqnarray}
e^{F_{1}}&=& \int D\hat{\chi}(\vec{r})D\sigma(\vec{r})
e^{-\int 
d\vec{r}\left(\frac{\beta\epsilon}{8\pi}\left|\vec{\nabla}\hat{\chi}(\vec{r})\right|^{2}+\frac{1}{2\lambda}
 (\sigma(\vec{r})-\beta 
pe\hat{\chi}(\vec{r}))^{2}+\frac{\beta^{2}e^{2}}{2}\left(c_{+}e^{\beta e\chi_{c}}
 +c_{-}e^{-\beta e\chi_{c}}\right)\hat{\chi}(\vec{r})^{2}\right)} \nonumber \\
 &\times&e^{-M\int d\vec{r}d\vec{r}\,^{\prime}\sigma(\vec{r})
G_{c}(\vec{r},\vec{r}\,^{\prime})\sigma(\vec{r}\,^{\prime})}
\end{eqnarray}
whence 
\begin{equation}
 F_{1}=-\frac{1}{2}\ln{{\rm Det} K}
\end{equation}
where $K$ is the kernel\\
\[ K=\left( \begin{array}{cc} 
-\frac{\beta\epsilon}{8\pi}\Delta+\frac{\beta^{2}p^{2}e^{2}}{2\lambda}+
\frac{\beta^{2}e^{2}}{2}
\left(c_{+}e^{\beta e\chi_{c}}+c_{-}e^{-\beta e\chi_{c}}\right)& 
-\frac{\beta pe}{2\lambda}\\
-\frac{\beta pe}{2\lambda} &
MG_{c}+\frac{1}{2\lambda} \end{array} \right) \]

 This determinant must be rendered well-defined by an explicit cutoff procedure, such as
 the lattice. On a lattice with N points, we then have the problem of evaluating a 2Nx2N
 determinant. If the polymer is restricted to a subregion of  N$_{p}$ points, the 
lower right hand  NxN block in $K$ is non-sparse only in a N$_{p}$xN$_{p}$ subblock of the  kernel $G_{c}$.  The evaluation of $G_{c}$ is facilitated by the observation that
\begin{equation}
 \sum_{n\neq 0}
\frac{\Psi_{n}(\vec{r})\Psi_{n}(\vec{r}\,^{\prime})}{E_{n}-E_{0}}=
\lim_{\eta\rightarrow 0} 
\left((H-E_{0})\frac{1}{(H-E_{0})^{2}+\eta^{2}}\right)(\vec{r},\vec{r}\,^{\prime})
\end{equation}
with the required inversion involving only a sparse matrix 
$(H-E_{0})^{2}+\eta^{2}$.

\newpage

\section{Mean-field results for spherically symmetric systems}

 For spherically symmetric systems, the extremization of the functional $F$ in 
(14) is much simplified. It is convenient to express all distances in units of the effective monomer
 size (Kuhn length) $a_{p}$, and to define rescaled radial functions $f(r)$, 
$g(r)$ as
 follows
\begin{eqnarray}
  f(r) &=& \beta e \chi_{c}(r) \label{eq:f} \\
  g(r) &=& r\Psi_{0}(r)
\end{eqnarray}
It is also convenient to rescale the auxiliary field $\omega_{c}$ to a 
 dimensionless one, $h(r)\equiv \omega_{c}a_{p}^{3}$. Introducing dimensionless parameters $\eta$, $\xi_{\pm},$ and $\zeta$ 
\begin{eqnarray}
  \eta &\equiv& \frac{\epsilon a_{p}}{2\beta e^{2}} \\
  \xi_{\pm} &\equiv& 4\pi c_{\pm}a_{p}^{3} \\
  \zeta &\equiv& 4\pi\frac{\lambda}{a_{p}^{3}}
\end{eqnarray}
 the saddle-point functional can be written as a one-dimensional integral
\begin{equation}
 F=\int \left\lbrace \eta \left(\frac{df}{dr}\right)^{2}+\frac{1}{2}\zeta h(r)^{2}
  +\xi_{+}e^{f}+\xi_{-}e^{-f} \right\rbrace r^{2}dr - ME_{0} 
\end{equation}
where the radial wavefunction $g(r)$ of the associated Schr\"odinger Hamiltonian 
(13)
 satisfies
\begin{equation}  
\frac{1}{6}\frac{d^{2}g}{dr^{2}}=\left(\frac{\zeta}{4\pi}h(r)+pf(r)-E_{0}
\right)g(r)
\end{equation}
 The rescaled activity coefficients $\xi_{\pm}$ must be constrained by the 
appropriately rescaled versions of  (33):
\begin{equation}
    \xi_{\pm} = \frac{n_{\pm}}{\int e^{\pm f}r^{2}dr}
\end{equation}

   We have devised the following efficient procedure for the minimization of $F$ 
in (58).
 The arguments of the preceding section establish the convexity of $F$, implying a
 unique minimum. After discretizing the $r$ variable, the rescaled electrostatic 
potential function
 $f(r)$ and polymer density function $h(r)$ become finite arrays which can be
 updated alternately by the following procedure, which at each step takes us 
closer
 to the unique global extremum of $F$:
\begin{enumerate}
\item Choose a reasonable starting value for the fields $f$, $h$.
\item  Define a new field $\sigma(r)\equiv \frac{\zeta}{4\pi}h(r)+pf(r)$, so that 
$E_0$ in (58)
 is the lowest eigenvalue of the operator 
$H=-\frac{1}{6}\frac{d^{2}}{dr^{2}}+\sigma(r)$.
  Once $r$ is discretized, $H$ becomes a tridiagonal matrix. For the rest of the 
calculation,
 we minimize with respect to $f(r)$ and $\sigma(r)$.
\item  Minimize $F$ with respect to $\sigma(r)$ for each discrete value of $r$. This
 requires a rapid calculation of $E_0$ as a function of $\sigma(r)$. The ground state eigenvalue of a tridiagonal matrix
 (indeed, any ordinally located eigenvalue) can be readily extracted
 by Sturm sequence methods \cite{Golub}, and the functional $F$ minimized quickly with respect
 to $\sigma(r)$ by golden section bracketing  \cite{NumRec}. The latter approach is
 foolproof as we have a strictly convex dependence on $\sigma(r)$ for all $r$.
\item Minimize $F$ with respect to $f(r)$ for each discrete value of $r$. This is trivial as the dependence of $F$ on $f(r)$ is explicit.
\item  Iterate until $F$ stabilizes at a minimum to some preassigned tolerance and/or the 
 saddle-point equations are satisfied to  a desired degree of accuracy.
\end{enumerate}
\begin{figure}[htp]
\hbox to \hsize{\hss\psfig{figure=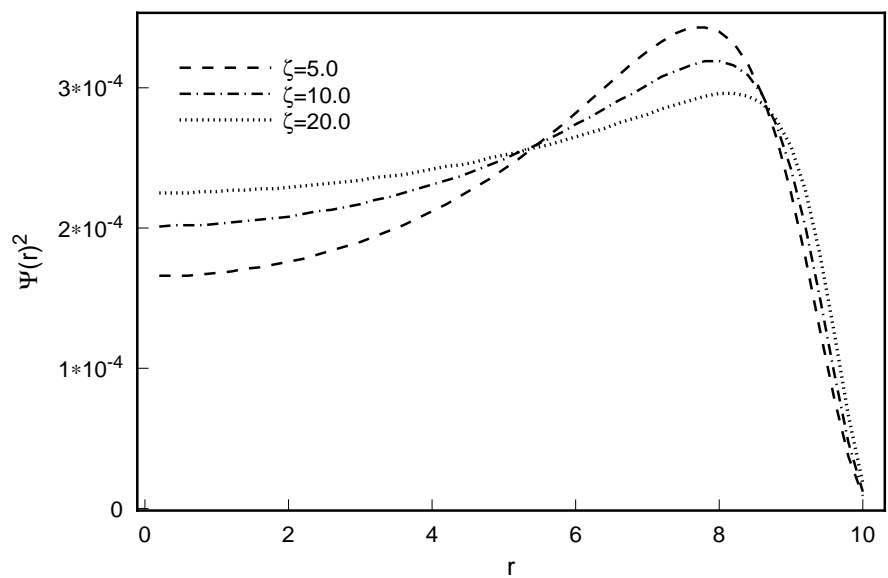,width=0.8\hsize}\hss}
\caption{ Monomer distribution for varying $\zeta$}
\end{figure}

\begin{figure}[htp]
\hbox to \hsize{\hss\psfig{figure=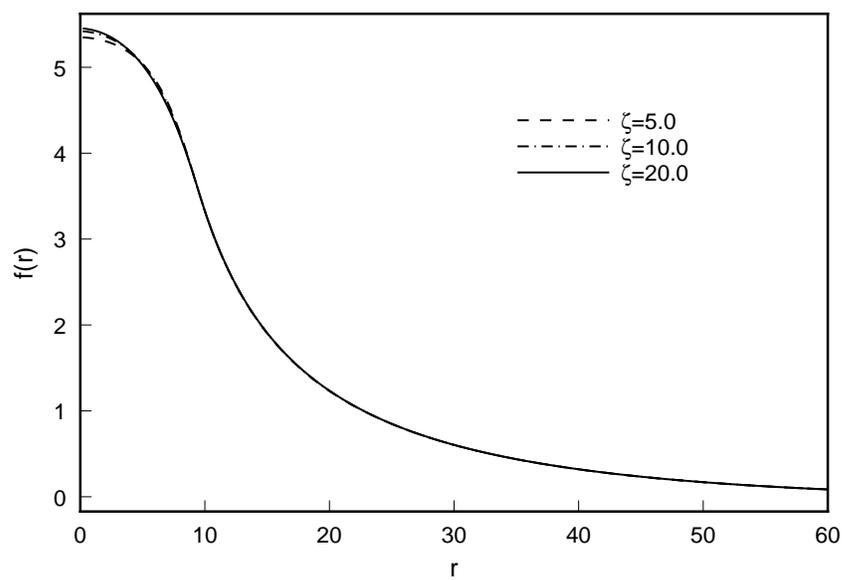,width=0.8\hsize}\hss}
\caption{ Electric Potential $f(r)$ for varying $\zeta$}
\end{figure}

   This algorithm was applied to a system consisting of a charged polymer with
 1000 monomer subunits trapped in a spherical cavity of radius 10$a_{p}$, with
 each monomer carrying a charge -0.1$e$. The ions ($n_{+}$=200 positive and
 $n_{-}$=100 negative ions) are free to move in a larger spherical region of 
radius
 100$a_{p}$. The parameter $\eta$ was taken to be unity. The results
 shown correspond to 1000 iterations, after which the free energy is stabilized to
 5 significant figures (such a run takes a few minutes on a 400 MHz Pentium 
processor).
 The effect of
 variation of the excluded volume parameter $\zeta$ on the monomer density
 $\Psi_{0}^{2}(r)$ is shown in Fig. (1). As for the case of the uncharged polymer, 
 increasing the excluded volume parameter results in a flattening of the 
distribution
 near the origin. However, in the presence of charge, we now find that (at least
 for the parameter range studied here) the electrostatic repulsion sufficiently 
counteracts the
 tendency of the monomers to crowd into the central region to produce a {\em dip} 
in the
 distribution for small $r$. The mean-field rescaled electric potential $f(r)$ is 
 essentially constant over this variation of $\zeta$,
 as shown in Fig. (2) \cite{pot}.  The exponential
 screening outside the cavity of radius 10$a_{p}$ confining the polymer is 
evident.

  The effect of varying monomer charge $p$ (with the excluded volume parameter
 $\zeta$ held fixed at 20.0) is illustrated in Figs. (3,4). In particular, as 
$p\rightarrow0$ we
 recover the distinctive flat behavior at small $r$ characteristic of uncharged polymers (Fig. (3)).
 In these plots, the negative ion number is held constant at 100, with the number of positive ions adjusted to give charge neutrality.

\begin{figure}[htp]
\hbox to \hsize{\hss\psfig{figure=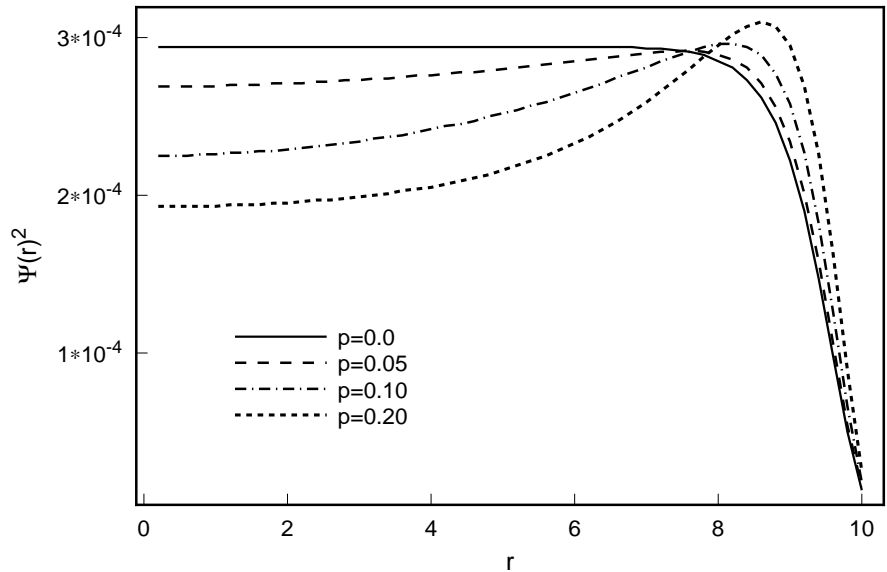,width=0.8\hsize}\hss}
\caption{ Monomer distribution for varying $p$}
\end{figure}

\begin{figure}[htp]
\hbox to \hsize{\hss\psfig{figure=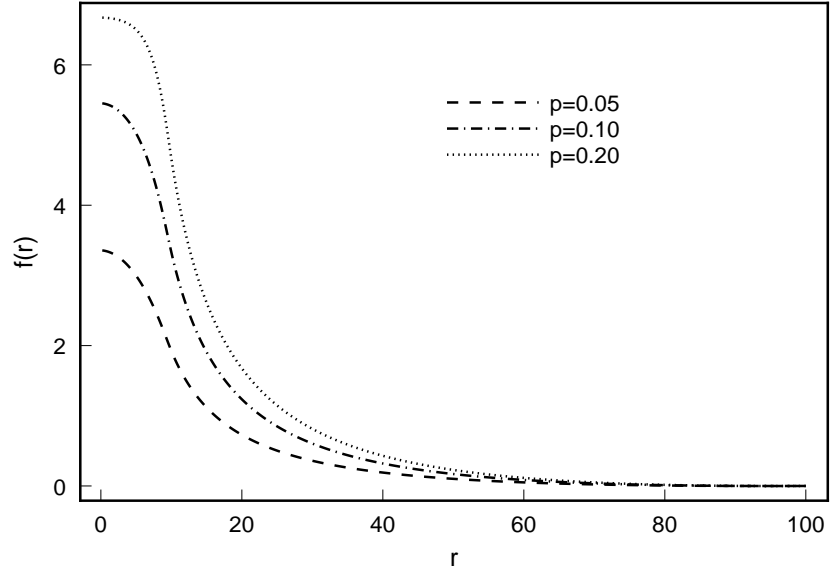,width=0.8\hsize}\hss}
\caption{ Electric Potential $f(r)$ for varying $p$}
\end{figure}
\vspace{3in}
\newpage

  Finally, we may also study the effect of varying salt concentrations. Increasing
 the background ion density results in higher screening of the 
 electrostatic potential, so the effect on the monomer distribution is similar to
 that obtained by varying the average monomer charge $p$. In Figs. (5,6)
 we show the monomer distribution and electrostatic potential $f(r)$ for fixed
 $\zeta$=10 and $p$=0.1, varying the number of negative mobile ions $n_{-}$,
 with $n_{+}=n_{-}+100$ for charge neutrality.
 
\begin{figure}[htp]
\hbox to \hsize{\hss\psfig{figure=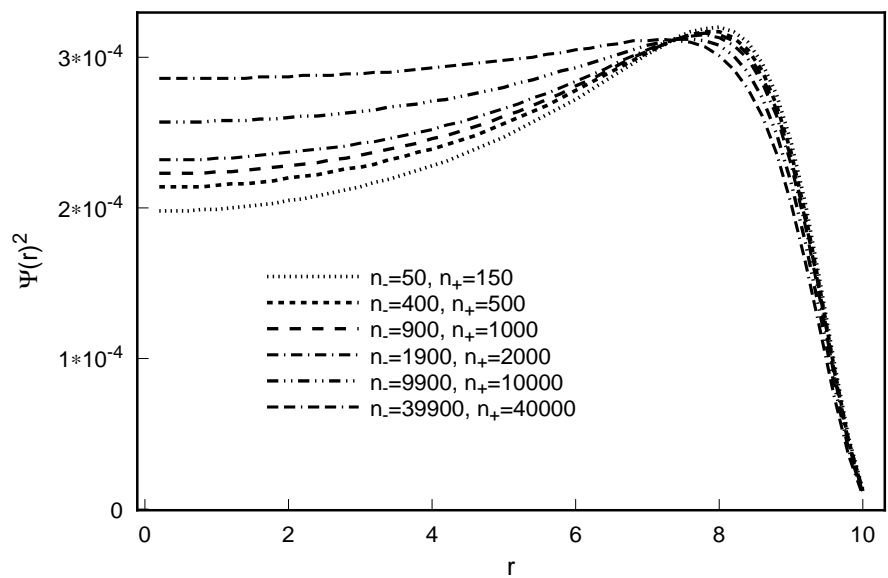,width=0.8\hsize}\hss}
\caption{ Monomer distribution for varying $n_{-}$}
\end{figure}

\begin{figure}[htp]
\hbox to \hsize{\hss\psfig{figure=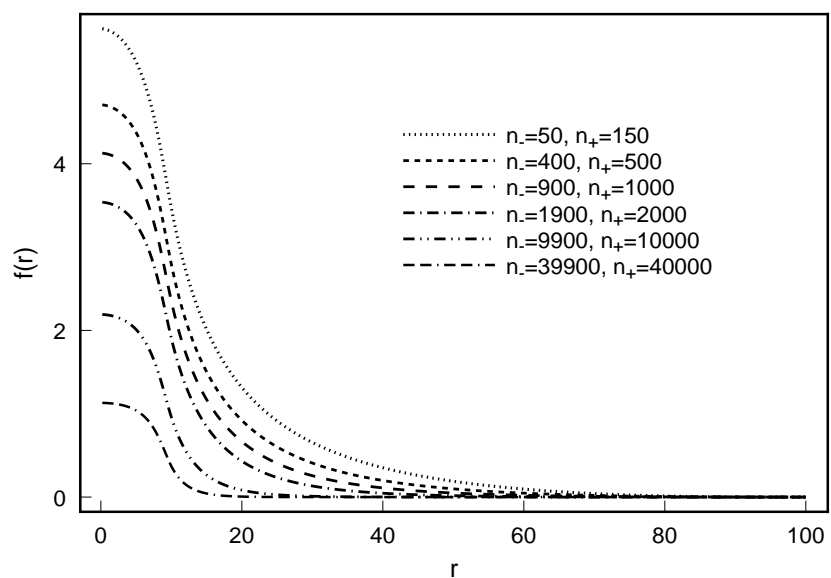,width=0.8\hsize}\hss}
\caption{ Electric Potential $f(r)$ for varying $n_{-}$}
\end{figure}
\vspace{3in}
\newpage

\section{Mean-field results using three-dimensional lattice field theory}

In the previous section we showed how the theory developed in
this paper can be applied to the problem of a charged polymer confined to a 
spherical cavity and immersed in an electrolyte solution. We now present the 
solution of the same problem using three-dimensional (3D) lattice field theory,
which does not hinge on special 
symmetry properties of the system, and thus illustrate
a numerical procedure for dealing with
systems of arbitrary shape and complexity. Colloidal
suspensions in polyelectrolyte solutions, which may be useful for
for a variety of technological applications,
such as new optical materials and devices
(e.g., narrow-band optical rejection filters, pump-probe laser apparati,
optical display panels, etc. \cite{Asher0,Asher01}), are systems of
this type.

	We will solve the discretized versions of equations (22) and (23) 
simultaneously on a 3D lattice. It is convenient to multiply equation (22) by 
$a_{l}^{3}$, where $a_{l}$ is the lattice spacing, and to rescale according to: 
$f(\vec{r}){\rightarrow}{\beta}e{\chi}_{c}(\vec{r})$, ${\psi}_{0}(\vec{r}) 
{\rightarrow} a_{l}^{3/2}\psi_{0}(\vec{r})$. Thus, all variables and parameters 
become dimensionless and the two discretized equations are:
\begin{equation}
\alpha\sum_{\vec{m}}\Delta_{\vec{n}\vec{m}}f_{\vec{m}}=\gamma_{+}e^{f_{\vec{n}}-V_{\vec{n}}}-\gamma_{-}e^{-f_{\vec{n}}-V_{\vec{n}}}-Mp\psi_{\vec{n}}^{2}
\end{equation}
\begin{equation}
\frac{a_{p}^{2}}{6a_{l}^{2}}\sum_{\vec{m}}\Delta_{\vec{n}\vec{m}}\psi_{\vec{m}}=\frac{{\lambda}M}{a_{l}^{3}}\psi_{\vec{n}}^{2}\psi_{\vec{n}}+pf_{\vec{n}}\psi_{\vec{n}}-E_{0}\psi_{\vec{n}}
\end{equation}
where
\begin{equation}
\alpha=\frac{{\varepsilon}a_{l}}{4\pi{\beta}e^{2}},
\end{equation}
\begin{equation}
\gamma_{\pm}=\frac{n_{\pm}}{\sum_{\vec{n}}e^{\pm f_{\vec{n}}}},
\end{equation}
and the wavefunction is dimensionless and normalized according to the rule
\begin{equation}
\sum_{\vec{n}}\psi_{\vec{n}}^{2}=1  \; 
\end{equation}

	We solve equations (61) and (62) simultaneously using
the following procedure. First, equation (62) is solved for $f_{\vec{n}}=0$
 and ignoring the non-linear (monomer repulsion) term. The resulting 
$\psi_{\vec{n}}$ (wave-function for a free particle in a sphere) is given to 
equation (61), which is solved at each lattice point by the Newton-Raphson method. 
The process is repeated and the coefficients $\gamma_{\pm}$ are updated after each 
iteration with the current field, until a predetermined desired accuracy is 
achieved. Then the resulting $f_{\vec{n}}$ is fed into equation (62) which is 
solved for a new $\psi_{\vec{n}}$ to be given to equation (61). Equation (62) is 
solved to a predetermined desired accuracy by the Lanczos approach, which is 
appropriate for a large sparse matrix like the one representing our Hamiltonian. This method of computation is very well-suited for implementation on massively parallel platforms which should make it possible to study even very large lattices with this approach.
From then on, the potential $f_{\vec{n}}$ to be given to (62) for the next 
iteration is updated slowly by adding a small fraction of the new $f_{\vec{n}}$ to 
the old one, obtained from the previous iteration. The same overrelaxation procedure is used for 
updating $\psi_{\vec{n}}^{2}$ in the non-linear term of the Schr\"odinger equation 
(62). Such a gradual iteration procedure is necessary in order to avoid
an unstable bifurcation
between two unphysical states (which is commonly encountered when
solving non-linear differential equations), and it converges to the simultaneous 
solution of the two equations. We have already shown that the functional in (14) 
has an unique minimum, the condition for which is given by the two equations, (61) 
and (62). Therefore, once we have converged to a solution of these two equations, 
we are guaranteed to have reached the unique mean-field solution of the problem.

	We have applied the procedure described above to a system of a negatively 
charged polymer of 1000 monomer units, confined to a spherical cavity of radius 
$20a_{p}$, which is immersed in an electrolyte solution confined to a larger 
sphere of radius $40a_{p}$. The Kuhn length $a_{p}$ has been chosen to be 5\AA. To 
illustrate the stability and accuracy of the procedure, we compare some of the 3D 
results with the ones obtained through the one-dimensional (1D) calculation of 
Section 6, which are practically exact. In Fig. (7) we show how the 3D results for 
$\psi_{0}^{2}(\vec{r})$ approach the exact 1D result as the number of the lattice 
points on the side of the 3D cube containing the system, L, is increased from 40 
to 60 and 80, for the following parameters: $\zeta=15$, $p=0.1$ and $n_{-}=6$, 
where $n_{-}$ is the number of the negative ions in the system, while the number 
of the positive ions is adjusted so that electroneutrality is preserved, and the 
relationship between $\zeta$ and $\lambda$ (in equation (62))
 is given in (57). 

%
%
\begin{figure}[htp]
\hbox to 
\hsize{\hss\psfig{figure=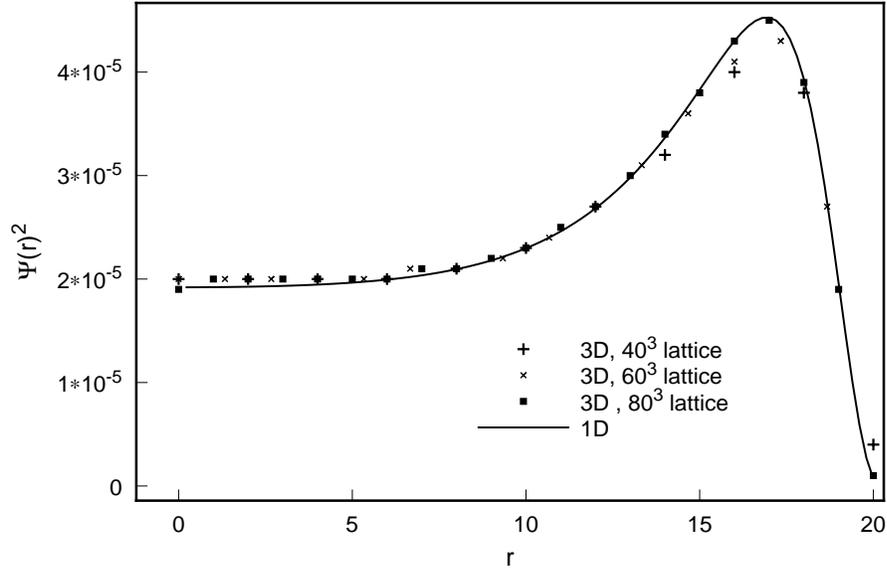,width=0.8\hsize}\hss}
\caption{ Monomer distribution for varying lattice size}
\end{figure}

	In Fig. (8) we show the effect of variation of $\zeta$ on the probability 
distribution $\psi_{0}^{2}(\vec{r})$. For comparison we have included two results 
obtained by the 1D method of Section 6. The results are analogous to the ones in 
Fig. (1). Similarly, the rescaled electrostatic potential $f(\vec{r})$ varies 
little with $\zeta$, Fig. (9).
\vspace{1in}

%
\begin{figure}[htp]
\hbox to \hsize{\hss\psfig{figure=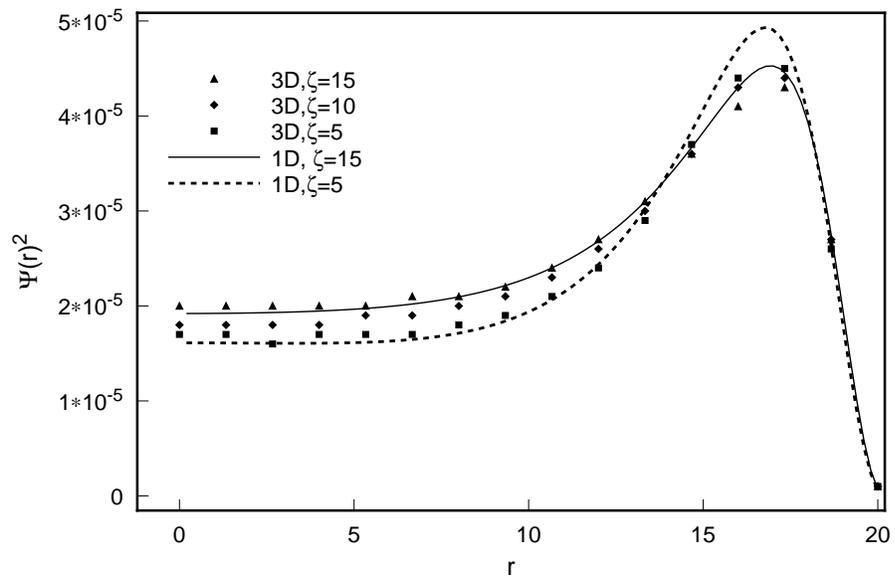,width=0.8\hsize}\hss}
\caption{ Monomer distribution for varying $\zeta$}
\end{figure}

%
%
\begin{figure}[htp]
\hbox to \hsize{\hss\psfig{figure=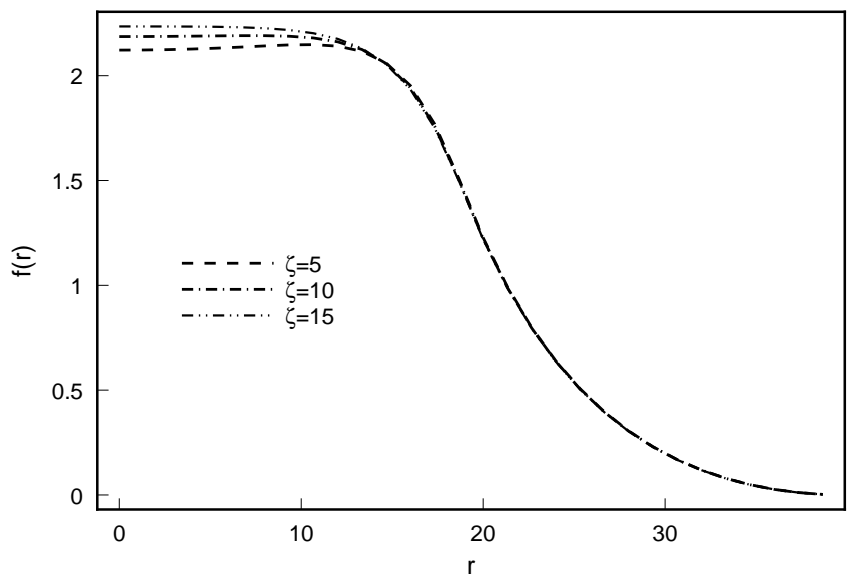,width=0.8\hsize}\hss}
\caption{ Electric Potential $f(r)$ for varying $\zeta$}
\end{figure}
\vspace{3in}
\newpage

	In Figs. (10) and (11) we present results for $\psi_{0}^{2}(\vec{r})$ and 
$f(\vec{r})$ for varying monomer charge $p$, fixing $\zeta=15$. In these plots the 
number of negative ions is $n_{-}=6$.

%
%
\begin{figure}[htp]
\hbox to \hsize{\hss\psfig{figure=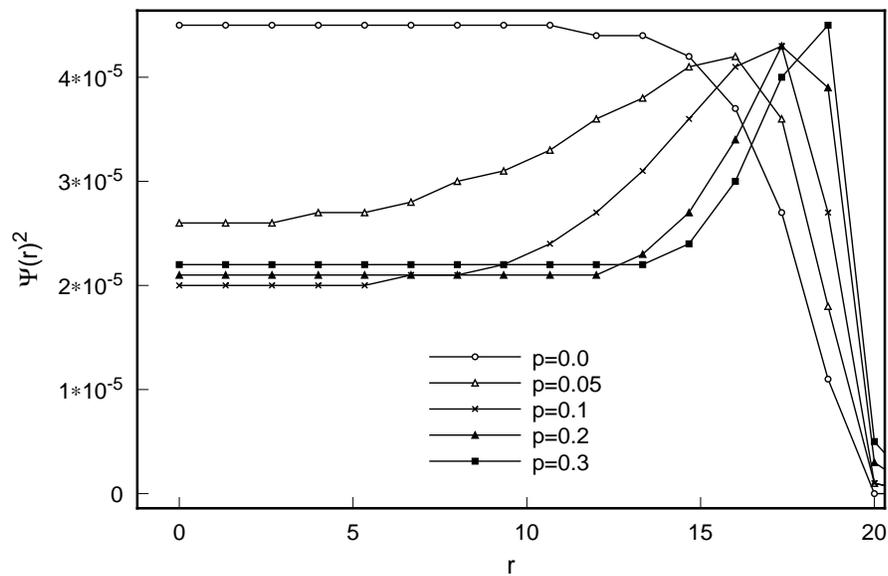,width=0.8\hsize}\hss}
\caption{ Monomer distribution for varying $p$}
\end{figure}

%
\begin{figure}[htp]
\hbox to \hsize{\hss\psfig{figure=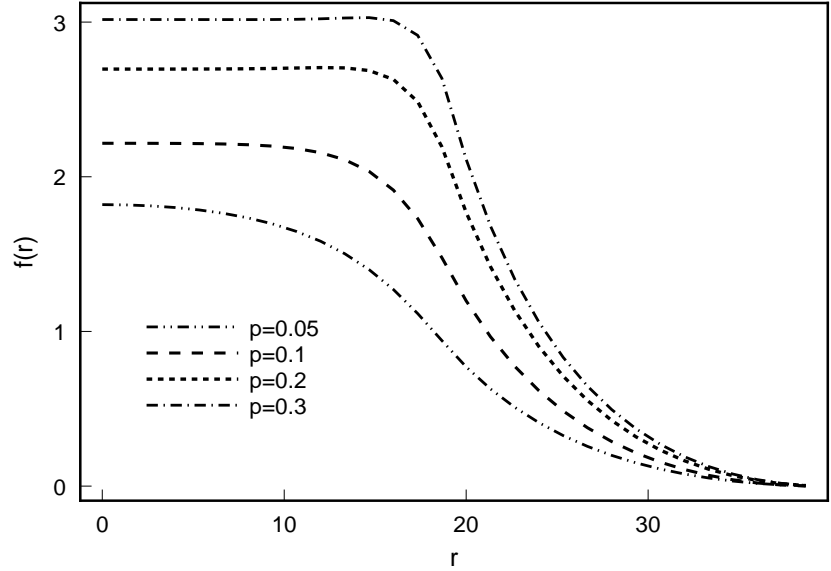,width=0.8\hsize}\hss}
\caption{ Electric Potential $f(r)$ for varying $p$}
\end{figure}
\vspace{3in}
\newpage

	And finally, in Figs. (12) and (13) we illustrate the effect of varying the 
number of impurity ions (concentrations) on the monomer distribution and the 
potential $f(\vec{r})$. Here, again, we compare the 3D results with the ones 
obtained by the method of Section 6. Clearly, the agreement between the two 
approaches is good.

%
\begin{figure}[htp]
\hbox to \hsize{\hss\psfig{figure=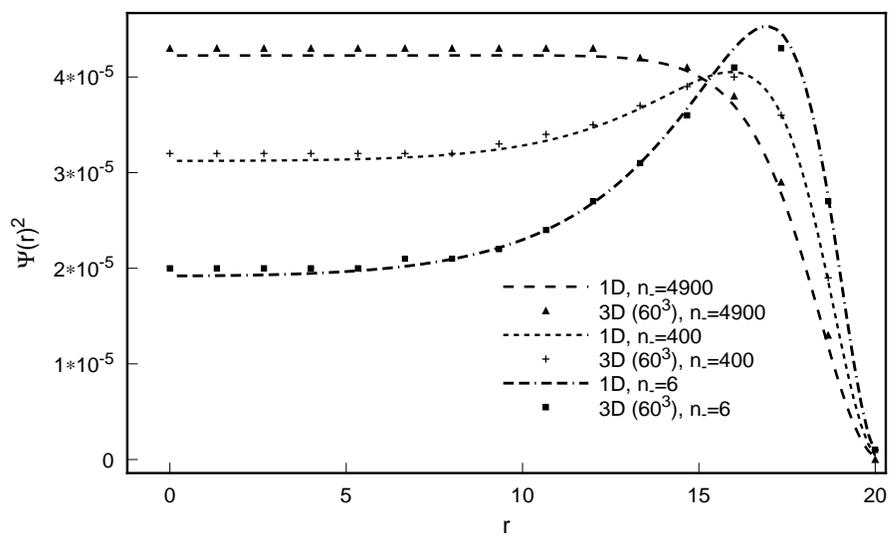,width=0.8\hsize}\hss}
\caption{ Monomer distribution for varying $n_{-}$}
\end{figure}

%
%
\begin{figure}[htp]
\hbox to \hsize{\hss\psfig{figure=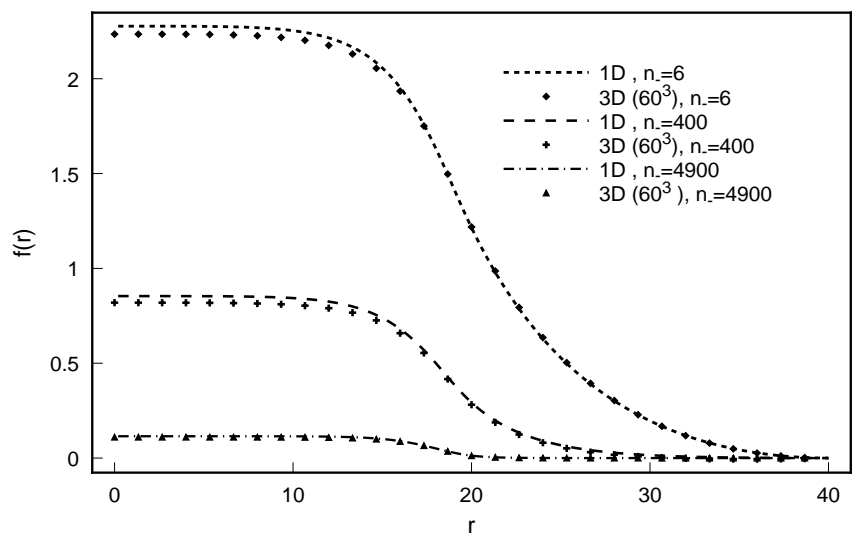,width=0.8\hsize}\hss}
\caption{ Electric Potential $f(r)$ for varying $n_{-}$}
\end{figure}
\vspace{7in}

\newpage

\section{Discussion}

	We have derived the mean-field equations for a coupled ionic-polymer 
system by doing a saddle-point analysis on a functional integral representing
the partition function of the system, (11).
This analysis shows that all mean-field level thermodynamical properties
are obtained by extremizing an appropriate
real-valued functional (14).
Moreover, we have shown that the functional (14) possesses a single minimum, 
guaranteeing a unique solution of the coupled mean-field equations. We have also 
described two different numerical procedures for finding the mean-field solution, 
and have applied those to the problem of a charged polymer confined to a spherical 
cavity and immersed in an electrolyte solution.

	Although our calculations were intended as an illustration of the 
advantages of the approach, they have given us some interesting insight into the 
physics of confined polyelectrolytes. It has been suggested that materials 
consisting of spherical voids imbedded in a polymer gel can be used as so called 
``entropic trapping devices,'' in which macromolecules such as polymers and DNA 
could be trapped and separated \cite{Asher,Asher2}. Therefore, such materials may 
have important applications in specific biochemical trapping, or even as 
microreactors for applications in organic, bioengineering and combinatorial 
synthesis \cite{Asher}.

	It has been hypothesized
\cite{Cas,CasTag,Baum,Mutu} that such trapping is a result of the higher 
conformational entropy ``enjoyed'' by the polymer in the large spherical void, as 
compared to the narrow channels connecting the voids within the gel. Our 
calculations reveal another important aspect of the trapping phenomenon---namely, 
from our results it becomes clear that electrostatic interactions also play a very 
important role in it. From Figs. (3), (5), (10) and (11) it is seen that polymers 
with
high monomer charge $p$, or in dilute electrolyte solution, have a distribution 
function with a peak near the edge of the spherical cavity, while polymers with 
low monomer charge or in concentrated electrolyte solution (where the monomers are 
highly screened by the impurity ions, thus experiencing weaker repulsion from each 
other), have a more flattened distribution, and are more likely to be found near 
the center, rather than the edge of the cavity.
In addition, highly charged monomers would
possess higher energy in the voids, due to the stronger repulsion from each other 
as they are brought closer together in the folded polymer. Therefore, we expect 
that polymers with relatively low average monomer charge or in a concentrated 
electrolyte solution would be easier to trap in spherical voids. Usually the 
monomer charge is approximately constant, and it is the impurity ion concentration 
that can be varied in the laboratory. We suggest that experiments be performed
to investigate polymer trapping dependence on electrolyte concentration.

	We also expect that the lattice field theory approach to the statistical 
mechanics of charged polymers in electrolyte solution which has been developed in the
present work 
will be useful for studying low symmetry systems involving complex liquids, such 
as colloidal suspensions in polyelectrolyte solutions.


\begin{thebibliography}{99}
\bibitem{footeps}
It was shown in \cite{LFT3} that an arbitrary dielectric
profile $\epsilon (\vec r)$ can be incorporated into the LFT formalism
in a straightforward way. For simplicity, we consider only the
prototypical case of constant $\epsilon$ here.
\bibitem{footmac}
It is also possible to include a charge density term that accounts
for one or many fixed macroions, such as the charged colloid 
particles studied in \cite{LFT1}--\cite{LFT5}.  We will not
consider this complication in the present work, because the system
upon which we focus, a charged polymer in an electrolyte solution,
is interesting in its own right.
\bibitem{footground} The condition of ground state dominance naturally
requires that the polymer be confined by a potential (e.g., walls of
a container), so that the energy spectrum of the Hamiltonian operator
in (\ref{eq:H}) is discrete and $M(E_1-E_0) \gg 1$, $E_1$ being
the energy eigenvalue corresponding to the first excited state.
\end{thebibliography}
\end{document}